\documentclass[11pt]{article}
\usepackage{moriond,epsfig}

\def\be{\begin{equation}}
\def\ee{\end{equation}}
\def\bea{\begin{eqnarray}}
\def\eea{\end{eqnarray}}

\def\lsim{\mathrel {\vcenter {\baselineskip 0pt \kern 0pt
    \hbox{$<$} \kern 0pt \hbox{$\sim$} }}}
\def\gsim{\mathrel {\vcenter {\baselineskip 0pt \kern 0pt
    \hbox{$>$} \kern 0pt \hbox{$\sim$} }}}
\begin{document}
\vspace*{4cm}
\title{$\Sigma^{+}\to p \mu^{+}\mu^{-}$:  Standard Model or New Particle?}

\author{ G. Valencia }

\address{Department of Physics, Iowa State University, Ames, IA 50011}

\maketitle\abstracts{
The HyperCP collaboration observed three events for the decay $\Sigma^+ \to p \mu^+ \mu^-$. They suggested that new physics may be required to understand the implied decay rate and the observed $m_{\mu\mu}$ distribution. Motivated by this result, we re-examine this mode. First within the standard model, and then assuming there is a new particle.
Within the SM we find that $\Sigma^+ \to p \mu^+ \mu^-$ is long-distance dominated and its rate falls within the range suggested by the HyperCP measurement. We then examine the conditions under which the observation is consistent with a light Higgs boson and find an explicit example that satisfies all the constraints: the light pseudoscalar Higgs boson in the next-to-minimal supersymmetric standard model (NMSSM).}

\section{Introduction}

The HyperCP collaboration has observed three events for the mode $\Sigma^+\to p \mu^+ \mu^-$ \cite{Park:2005ek}. A striking feature of the result is that the three events have the same muon pair invariant mass, 214.3 MeV. HyperCP estimates the probability for this clustering at $0.8\%$ using a ``form factor'' distribution for the standard model expectations \cite{Bergstrom:1987wr}. 

This observation invites two calculations and we report on the results in this talk. First we present the best possible prediction for the Standard Model expectation. Since there are no known particles of mass 214 MeV, we do not expect a peak at that muon pair invariant mass. However, we need to know whether the SM distribution is narrower or wider than the form used by HyperCP to assess the significance of the clustering. Even if the three events represent new physics, it is necessary to know the SM level  in order to determine if HyperCP should have seen events at other values of $m_{\mu\mu}$.

The second calculation involves assuming that the observed events are indeed evidence for a new particle and confronting this observation with existing constraints from kaon and B physics. 
In particular we study the conditions under which the observation is consistent with a light Higgs boson and find an explicit candidate for the new particle: the lightest CP-odd Higgs boson in the NMSSM, the $A_1^0$.
 
\section{Standard Model Calculation}

We first present the ingredients that enter the calculation within the SM \cite{He:2005yn}. The short distance contribution is too small to explain these events by four orders of magnitude, this decay is  long distance dominated as is the case in similar kaon modes.

The long distance contributions to $\Sigma^{+}\to p \mu^{+}\mu^{-}$ can be pictured schematically as arising from the $\Sigma^{+}\to p \gamma^\star$ process. There are four independent form factors allowed by electromagnetic gauge invariance,
\begin{eqnarray}   \label{M_BBg}
 {\cal M}(B_i\to B_f\gamma^*)& =
&- e G_{F}^{}\, \bar{B}_f^{} \left[ i\sigma^{\mu\nu}q_\mu^{}(a+b\gamma_5^{})
+(q^2\gamma^\nu-q^\nu\!\!\not{\!q}) (c+d\gamma_5^{}) \right] B_i^{}\, \varepsilon_\nu^{}   \,\,.
\end{eqnarray}
Two of the form factors, $a(q^2)$ and $c(q^2)$, are parity conserving 
whereas $b(q^2)$ and $d(q^2)$ are parity violating. In addition, two of the form factors are non-zero at $q^2=0$ and contribute to the radiative decay $\Sigma^+\to p\gamma$: $a(0)$ and $b(0)$. All four form factors are complex and receive imaginary parts from $N\pi$ intermediate states.

We estimate these imaginary parts by taking the weak vertex $\Sigma^+ \to N\pi$ from experiment  and using 
the $\,N\pi\to p\gamma^*\,$ scattering at lowest order in $\chi$PT (both conventional and heavy baryon). We check that our calculations agree with the existing ones at $q^2=0$.

To estimate the real part of the form factors we use $a(0)$ and $b(0)$, as determined from the width and decay distribution of the radiative decay  $\Sigma \to p \gamma$  up to a discrete ambiguity. We then assume that value for the range of $q^2$ needed. This is consistent with our finding that the imaginary parts of the form factors are smooth and slowly varying over the $q^2$ range of interest. 
Finally, the real parts of $c(q^2)$ and $d(q^2)$ are obtained using a vector meson dominance model.

There is some uncertainty in the calculation, but the resulting range, $1.6 \times 10^{-8}
\leq \ \ {\cal B}(\Sigma^{+}\to p \mu^{+}\mu^{-})_{SM}\ \ \leq
9.0\times 10^{-8}$,  is in good agreement with the measured rate, 
${\cal B}(\Sigma^{+}\to p \mu^{+}\mu^{-})=(8.6^{+6.6}_{-5.4}\pm5.5)\times10^{-8}$ \cite{Park:2005ek}. 
The predicted $m_{\mu\mu}$ distribution shows no peaks near 214~MeV (or elsewhere) and is slightly flatter than the form factor  used by HyperCP. This leads us to conclude that the probability of having the three events at the same invariant mass is about $0.5 \%$. Furthermore, the lower end of the range predicted for the rate is consistent with no events for HyperCP, allowing for the possibility of all three events being consistent with new physics.

\section{A new Particle with mass 214~MeV?}

We now turn to the interpretation of  the 3 HyperCP events as a new particle  \cite{Park:2005ek} with $M_{P^0}= 214.3~MeV$ and ${\cal B}(\Sigma^{+}\to p \mu^{+}\mu^{-})_{P^0} = 
(3.1^{+2.4}_{-1.9}\pm 1.5)\times 10^{-8} 
$.
The observation implies that this hypothetical new light state, $P^0$, is short lived, does not interact strongly, is narrow and decays only into $\mu^+\mu^-$, $e^+e^-$ or $\gamma\gamma$, and has a $\Delta S =1$, $\Delta I = 1/2$ coupling to $\bar{s} d$ quarks. There are three  questions to be answered and we address them in order. Why hasn't it been seen before? Is there a candidate for such a state? Where else could it be observed? 

\subsection{Why hasn't it been seen before?}

The most stringent constraint on a possible new particle $P^0$ is its non-observation in kaon decay. After all, the modes $K \to \pi \mu^+ \mu^-$ proceed via the same quark level transition as $\Sigma^+ \to p \mu^+ \mu^-$: $s\to d\mu^+ \mu^-$. Of the three experiments that have studied these modes: BNL865 \cite{Ma:1999uj}, HyperCP \cite{Park:2001cv} and NA48 \cite{Batley:2004wg} the one with most statistics was BNL865 \cite{Ma:1999uj} with 430 events, 30 of which were in their lowest bin $2m_\mu \lsim m_{\mu\mu} \lsim 225$~MeV where the signal would have been observed. Their observation shows no peaks in the $m_{\mu\mu}$ distribution, which is consistent with long distance SM physics. On that basis, the most optimistic scenario for the new physics hypothesis is to assume that all the 30 events in the first bin were due to $P^0$ which leads to a 95\% confidence limit bound ${\cal B}(K^+ \to \pi^+ P^0)\leq 8.7 \times 10^{-9}$ \cite{He:2006uu} (assuming that statistical errors dominate). This translates into a rate for $\Sigma^+ \to p P^0$ some 25 times too small to explain the HyperCP events. Similar results are obtained from the other kaon experiments, none of which saw a peak in their $m_{\mu\mu}$ distribution.

Another constraint arises from the non-observation of the hypothetical new particle in $b \to s \mu^+ \mu^-$. In this case both Belle and BaBar \cite{bbounds} have results that can be interpreted as a 95\% confidence level bound \cite{He:2006uu} ${\cal B}(B\to X_s P^0) \leq 8 \times 10^{-8}$.

\begin{figure}
\center{\psfig{figure=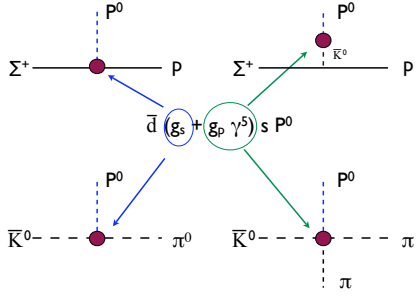,height=1.5in}}
\caption{$sd$ FCNC at the quark level: a scalar coupling only affects $K\to \pi P^0$ and a pseudoscalar coupling only affects $K\to \pi \pi P^0$. However, both affect $\Sigma \to p P^0$.
\label{sketch}}
\end{figure}

In Figure~\ref{sketch}, we can see schematically how it is possible for the new state to be observed in $\Sigma$ decay while not in $K^+$ decay: the kaon decay modes with only one pion in the final state only constrain the effective $|\Delta S| =1$ scalar coupling of the new state whereas the $\Sigma$ decay is sensitive also to the effective $|\Delta S| =1$ pseudoscalar coupling. Any viable model for $P^0$ will then have an effective scalar coupling about 25 times smaller than the corresponding pseudoscalar coupling \cite{He:2005we}.

In a similar manner, the constraints from $B$ decay require that the effective $bs$ coupling of $P^0$ be about an order of magnitude smaller than the corresponding $sd$ coupling scaled by $m_b/m_s$ and $(V_{ts}V_{tb}^\star) /(V_{ts}V_{td}^\star)$. The latter scaling is the appropriate one for one-loop Higgs penguins dominated by a top-quark and a $W$ boson in the intermediate state. A successful model for $P^0$ can not have these penguin diagrams dominating the effective FCNC of $P^0$ to down-type quarks.

We have also considered additional processes that can, in principle, constrain the interactions of the hypothetical $P^0$. $K-\bar{K}$ mixing allows an effective pseudoscalar coupling up to 50 times as large as required to explain the 3 HyperCP events. $K_L\to \mu^+\mu^-$ combined with the muon $g-2$ allow an effective pseudoscalar coupling as large as required. The muon $g-2$ allows a $P^0$ coupling to muons $g_{P\mu} \lsim 5 \times 10^{-4}$ which interestingly is about $m_\mu/v$ \cite{He:2005we,others}. 

\subsection{Is there a candidate for $P^0$?}

The possibility that $P^0$ is a light sgoldstino has been explored to some extent in the literature \cite{sgold}. Here, we pursue the possibility that $P^0$ is a light Higgs boson. 
For detailed phenomenology of kaon and hyperon decays involving a light Higgs particle it is necessary to recall that there are two types of contributions that are generally of similar size \cite{He:2006uu}. There are two-quark ``Higgs penguin'' contributions that arise at one loop order and depend on the details of the flavor changing sector of the model. There are also ``four-quark'' contributions arising from a tree-level, SM $W$ mediated $|\Delta S| =1$ decay, in which the light Higgs is radiated from any of the $u,d,s$ quarks or the $W$ boson via the tree-level flavor diagonal couplings of the Higgs. Both of these contributions can be calculated in chiral perturbation theory \cite{lighth}, and we do so at leading order.
Given our discussion in the previous section we concentrate on CP-odd or pseudoscalar Higgs bosons. 

One possible candidate for $P^0$ is the $A_1^0$ of the NMSSM. The Higgs sector of the NMSSM contains the usual two Higgs doublets $H_D$ and $H_U$ that appear in the MSSM plus the Higgs singlet N. In the physical spectrum there are two CP-odd scalars, of which the $A_1^0$ is the lightest. It has been proposed in the literature that this $A_1^0$ can be naturally light due to a global $U(1)$ symmetry \cite{Dobrescu:2000yn}.

The main features of the couplings of the $A_1^0$ to SM fields are as follows. Its coupling to $Zh$ ($h$ being the lightest CP even Higgs) is suppressed by $\tan\beta$ with respect to the MSSM $ZhA$ coupling allowing an evasion of LEP bounds in the large $\tan\beta$ regime. Its couplings to quarks are also suppressed by $\tan\beta$ with respect to those of the $A$ in the MSSM. This results, for large $\tan\beta$, in negligible couplings to up-type quarks. The couplings to down-type quarks are independent of $\tan\beta$ and can be written in terms of one parameter, $l_d$, which can be of order one \cite{Hiller:2004ii}: ${\cal L} =-l_d^{} m_d^{}\,\bar d\gamma_5^{}d(i A^0_1)/v 
-l_d^{} m_\ell^{}\,\bar \ell\gamma_5^{}\ell(i A^0_1)/v +\cdots $.

The four-quark contributions to $A_1^0$ production in light meson and hyperon decay are thus proportional to $l_d$ and independent of other parameters in the model. It is then straightforward to compute these contributions to the HyperCP case. We find \cite{He:2006fr}, 
${\cal B}_{4q}(\Sigma^+\to p A_1^0) = 1.7 \times 10^{-7} |l_d|^2$, 
which matches the central value of the HyperCP result for $l_d\sim 0.4$. The bad news is that this then leads to
${\cal B}_{4q}(K^+\to \pi^+ A_1^0) \sim  10^{-6}$, 
two orders of magnitude larger than the limit from BNL E865. The conclusion illustrated by this calculation is that it is relatively easy to have a light Higgs that matches the HyperCP observation but it is very hard to avoid seeing it in kaon decay as well.

However, there are also the two-quark contributions to the amplitudes and it is possible to arrange a cancellation between amplitudes that satisfies the kaon bounds. The two-quark contributions are much more model dependent than the four-quark contributions, but also suffer from additional constraints due to non-observation of $P^0$ in $B$ decay. We have not performed a full parameter scan, but rather illustrated that it is possible to satisfy all constraints. To this effect we start with the specific model considered  by Hiller \cite{Hiller:2004ii} and modify it accordingly. To suppress the FCNC in $B$ decay we consider $m_{\tilde{t}}=m_{\tilde{c}}$ and negligible squark mixing. The strength of the two-quark contribution to kaon decay is then tuned with $m_{\tilde{u}}-m_{\tilde{c}}$. We further consider (large) $\tan\beta = 30$, 
$m_{\tilde{t}}\sim 2.5$~TeV and $-\lambda x =150$~GeV to obtain neutralino masses in the 100-1500~GeV range \cite{He:2006fr}.
\begin{figure}
\center{\psfig{figure=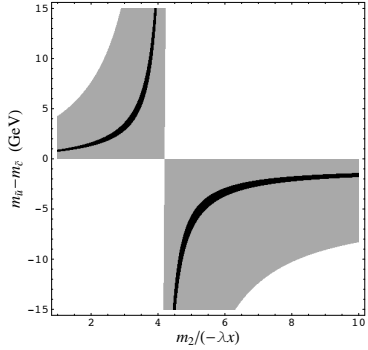,height=1.8in}}
\caption{Parameter space for  \,$m_{\tilde u}$$-$$m_{\tilde c}$\,  and  $m_2^{}/(-\lambda x)$  where
$A_1^0$ can explain the HyperCP events (gray regions) and simultaneously satisfy the kaon
bounds (black regions). The horizontal axis corresponds to parameters in the chargino mass matrix.}.
\label{fig:result}
\end{figure}
In Figure~\ref{fig:result} we show our results \cite{He:2006fr}: the light shaded region corresponds to parameters that reproduce the HyperCP observation. The dark shaded region corresponds to those points that also satisfy the kaon bounds. As mentioned before the overlapping region is significantly smaller due to the cancellation required to satisfy the kaon bounds.

\subsection{Where else can $P^0$ be observed?}

Finally,  we  explore other processes that can test the new particle hypothesis for the HyperCP result. We begin by considering only the effect of two-quark operators, assuming that the existing kaon bounds are bypassed because the effective $sd$ coupling is pseudoscalar. In this case the new state would show up in kaon decay modes with two pions in the final state and we can easily derive from the HyperCP measurement that (the errors reflect the experimental error only) \cite{He:2005we}
\begin{eqnarray}
{\cal B}(K_L \to \pi^+\pi^- P^0) &\approx & (1.8^{+1.6}_{-1.4})\times 10^{-9} \nonumber \\
{\cal B}(K_L \to \pi^0\pi^0 P^0) &\approx & (8.3^{+7.5}_{-6.6})\times 10^{-9}.
\label{q2res}
\end{eqnarray}
Both of these represent very significant enhancements over the corresponding SM rates and may be accessible to KTeV or NA48. In a similar manner this scenario results in \cite{He:2005we,Deshpande:2005mb}
\begin{eqnarray}
{\cal B}(\Omega^- \to \Xi^- P^0) &\approx & (2.0^{+1.6}_{-1.2})\times 10^{-6}.
\end{eqnarray}
The best upper bound for this mode, also from HyperCP \cite{solomey}, is $6.1 \times 10^{-6}$.

If the new state $P^0$ is a light Higgs, then there are other processes that are sensitive only to its flavor diagonal couplings \cite{Prades:1990vn} (or four-quark operators). For example  the modes $V \to \gamma A_1^0$ have been proposed in the literature \cite{Mangano:2007gi}. The results are  that ${\cal B}(\Upsilon_{1S}\to \gamma A_1^0)$ can reach about $1 \times 10^{-4}  l_d^2$ and may be accessible to the B factories. Similarly ${\cal B}(\phi\to \gamma A_1^0$ can reach $1.4\times 10^{-8}  l_d^2$ and may be accessible to DA$\Phi$NE \cite{Mangano:2007gi}. In a similar spirit we have proposed the modes $\eta \to \pi \pi A_1^0$ where we can predict \cite{He:2008zw} ${\cal B}(\eta \to \pi^+ \pi^- A_1^0) = 5.4 \times 10^{-7}l_d^2$, again possibly accessible to DA$\Phi$NE.

\begin{figure}
\center{\psfig{figure=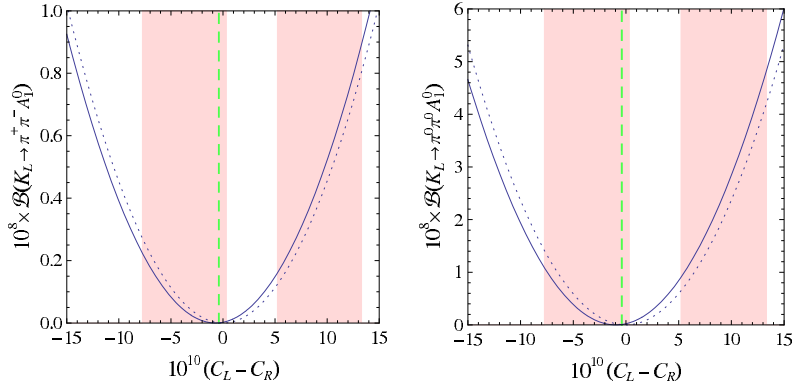,height=2.0in}}
\caption{Predicted branching ratios (solid curves) for  \,$K_L^{}\to\pi^+\pi^-A_1^0$\,
and  \,$K_L^{}\to\pi^0\pi^0A_1^0$\, with \,$l_d^{}=0.35$\,. The horizontal axis corresponds to the size of $g_P$.}
\label{fig:fullresult}
\end{figure}

When the four-quark contributions are added to the two-quark contributions in the NMSSM (using parameters as in Hiller \cite{Hiller:2004ii} and Xiandong \cite{Xiangdong:2007vv}) the results of Eq.~\ref{q2res} are modified. An example of the resulting predictions for the rate of the kaon modes is shown in Fig.~\ref{fig:fullresult}. Full details can be found in the paper \cite{He:2008zw}, but the $x$-axis is related to the strength of the two-quark contribution though an effective $g_P$ and the strength of the four-quark contribution is kept fixed. The dotted curves result from the two-quark contributions alone.
The shaded (pink) bands indicate the allowed ranges of \,$C_L^{}$$-$$C_R^{}$\, when the two and four-quark contributions have the same sign \cite{He:2008zw}. 
Each vertical (green) dashed line corresponds to the special 
case \cite{He:2006fr} of chargino dominated penguins.

\section{Conclusions}

The decay $\Sigma^+ \to p \mu^+\mu^-$ within the SM is long distance dominated and the predicted rate is in the right range to explain the HyperCP observation. However, the predicted $m_{\mu\mu}$ distribution makes it unlikely to find the three events at the same mass ($P\lsim 0.8\%$). Existing constraints from kaon and B physics allow a new particle interpretation of the HyperCP result provided that the FCNC couplings of the new particle are mostly pseudoscalar and smaller for $b\to s$ transitions  than naive scaling with CKM angles would predict.

The NMSSM has a CP-odd Higgs boson, the $A_1^0$ that could be as light as the required 214~MeV. Its diagonal couplings to quarks and muons in the large $\tan\beta$ limit can have the right size as well. There are several modes that can test this hypothesis independently from the details of the flavor changing sector of the model: $\Upsilon_{1S}\to \gamma A_1^0$, $\phi\to \gamma A_1^0$ and $\eta \to \pi\pi A_1^0$. 

It is harder to suppress the effective scalar $sd$ coupling that appears in this model to the level required to satisfy the existing kaon bounds, but it is possible for certain values of the relevant parameters. The measurement of one of the modes $K_L \to \pi \pi \mu^+ \mu^-$ can confirm or refute this scenario.

\section*{Acknowledgments}

This work was done in collaboration with Jusak Tandean and Xiao-Gang He. It was supported in part by DOE under contract 
number DE-FG02-01ER41155.

\end{document}